\documentclass[12pt,a4paper]{article}

\usepackage{geometry}
\usepackage{verbatim}
\usepackage{graphicx}
\usepackage{subfigure}
\usepackage{float}
\usepackage{bm}
\usepackage{txfonts}
\usepackage{array}
\usepackage{cite}
\usepackage{CJK,CJKnumb,CJKulem}
\usepackage{color}
\usepackage{url}

\newcommand{\alfven}{Alfv\'en\ }

\begin{document}
\begin{center}
\Large\textbf{Mode structure symmetry breaking of reversed shear Alfv\'en eigenmodes and its impact on the generation of parallel velocity asymmetries in energetic particle distribution}
\end{center}

Guo Meng $^{1}$, Philipp Lauber$^{1}$, Xin Wang $^{1}$ and Zhixin Lu $^{1}$
  \\
$^{1}$Max-Planck-Institut f\"ur Plasmaphysik,  85748 Garching, Germany \\

\date{\today}
\begin{abstract}
In this work, the gyrokinetic eigenvalue code LIGKA, the drift-kinetic/MHD hybrid code HMGC and the  gyrokinetic full-f code TRIMEG-GKX are employed to study the mode structure details of Reversed Shear Alfv\'en Eigenmodes (RSAEs). 
Using the parameters from an ASDEX-Upgrade plasma, a benchmark with the three different physical models for RSAE without and with Energetic Particles (EPs) is carried out. 
Reasonable agreement has been found for the mode frequency and the growth rate. 
Mode structure symmetry breaking (MSSB) is observed when EPs are included, due to  the EPs' non-perturbative effects. It is found that the MSSB properties are featured by a finite radial wave phase velocity, and the linear mode structure can be well described by an analytical complex Gaussian expression $A(s)=e^{- \sigma (s-s_0)^2}$ with complex parameters $\sigma$ and $s_0$, where $s$ is the normalized radial coordinate.
The mode structure is distorted in opposite {manners} when the EP drive shifted from one side of $q_{min}$ to the other side, and specifically, a non-zero average radial wave number $\langle k_s\rangle$ with opposite signs is generated.
The initial EP density profiles and the corresponding mode structures have been used as the input of HAGIS code to study the EP transport. The parallel velocity of EPs is generated in opposite directions, due to different values of the average radial wave number $\langle k_s\rangle$, corresponding to different initial EP density profiles with EP drive shifted away from the $q_{min}$.




(Some figures may appear in colour only in the online journal)


\end{abstract}

\section{\label{Introduction}Introduction}

Alfv\'en eigenmodes can be excited by the energetic particles (EPs) in tokamak plasmas and may limit the achievable EP concentration in the plasma core due to radial transport. Various theoretical and numerical methods have been developed for the modelling of EP transport \cite{nikolaiRBQnf,zonca14multi-scale,podesta14kick,bass2017nonlinear}. Previous perturbative models calculate the linear mode structure without considering EP kinetic effects and give a good estimate of the linear growth rate \cite{chen1984fishbone} and saturation level when the growth rate is much smaller than the real frequency \cite{pinches1998hagis}. The EPs' non-perturbative effects on Alfv\'en mode broadening has been observed from gyrokinetic simulations of EP driven Toroidicity induced Alfv\'en eigenmodes (TAEs) \cite{wangGTC-SB2013PRL}, and thus, an enhanced EP transport was observed.
The ``distortion'' of the 2D mode structure has been also observed experimentally by Electron Cyclotron Emission Imaging (ECEI) \cite{tobias-SB-2011PRL,Classen-2010,ClassenLauber-1264}, and the 2D ECEI data has been used to compare with simulation results to identify the Alfv\'en eigenmodes excitation and to verify the various codes with experiments \cite{Sam19GTCbenchmark}.

For identifying the origin and the consequence of the mode structure distortion, recent work shows that the up-down asymmetry of the 2D mode structure can be induced by EP’s non-perturbative effect \cite{ma2015global,Lu-SB-18-NF}, and can change the EP transport properties, especially the parallel velocity profiles \cite{meng2020effects}. 
As shown in our previous work \cite{Lu-SB-18-NF,meng2020effects}, a complex Gaussian representation of the mode structure $A(s)=e^{- \sigma (s-s_0)^2}$ with complex parameters $\sigma$ and $s_0$ is adopted to describe the non-perturbative symmetry breaking features, where $s$ is the normalized radial coordinate.
The mode structure symmetry breaking (MSSB) of the Reversed Shear Alfv\'en Eigenmodes (RSAE), especially imaginary part of the parameter $s_0$, has an important effect on EPs parallel velocity profiles \cite{meng2020effects}. The complex radial wave vector is $k_s(s)\equiv-i\; d(\ln \hat\Phi(s))/ds$, whose real part represents the wave propagation in radial direction. By employing gyrokinetic simulations, it is shown that the mode location of the Toroidal Alfvén Eigenmode (TAE) driven by EPs depends on the EP gradient location \cite{wangGTC-SB2013PRL}. Nevertheless, the MSSB of beta induced Alfv\'en eigenmode is related to the distance between the location of the maximal EP drive (i.e. the maximum radial gradient) and the rational surface as shown in simulations and analytical studies \cite{Lu-SB-18-NF}. These studies show that the \alfven mode location and propagation both change due to the EPs' non-perturbative contributions.

In this work, RSAEs driven by EPs and the related EP transport are studied by hybrid and gyrokinetic codes based on ASDEX-Upgrade parameters. The gyrokinetic full-f code TRIMEG-GKX \cite{lu19trimeg,lu2021trimeg-gkx}, gyrokinetic/MHD hybrid code HMGC \cite{Briguglio-95-HMGC,HMGC1998} and gyrokinetic eigenvalue code LIGKA \cite{lauber2007ligka} are employed to identify the quantitative connection between the MSSB and the EP profiles such as the maximum EP drive location. 
This paper is organized as follows. In section \ref{sec2}, we list the physical models and the parameters used in the simulations.
In section \ref{sec3}, the RSAE simulations without EPs have been benchmarked by the three codes. Then, the RSAE symmetry breaking due to non-perturbative EP effects has been compared by the three codes and quantitative parameters describing the mode symmetry breaking properties are fitted. The EP transport due to the interaction with RSAE has been studied using HAGIS \cite{pinches1998hagis} with the more consistent mode structure and EP profiles. In section \ref{sec4}, a summary and an outlook for EP driven MSSB work are given.


\section{Simulation models and parameters}\label{sec2}
In this work, the EP non-perturbative effects on mode structures and EP transport are studied hierarchically.
While the EP-AE interaction takes place as a resonance phenomenon that requires a kinetic treatment of the EPs, the bulk plasma can be treated with different levels of approximations. The gyrokinetic codes are closest to first principles with the most important physics included, but require long simulation times and consume significant amount of computing resources, especially when studying the nonlinear, multiple spatial and temporal scale physics \cite{wang2010nonlinear,wangGTC-SB2013PRL,lu19trimeg}.
A variety of codes made specific simplifications to achieve the balance between the computational cost and important physics. Hybrid codes such as HMGC \cite{Briguglio-95-HMGC,HMGC1998} and MEGA \cite{MEGA1998} treat background plasma as MHD and EPs as a kinetic species. This simplification allows the simulations to tackle long time scale physics such as abrupt massive loss of EPs \cite{bierwage2018natrue-com}. As a further simplification, the Alfv\'en mode structure and frequency can be taken as fixed and only the wave-particle interaction is kept for studies of the saturation and transport levels of EP driven Alfv\'en modes \cite{pinches1998hagis}. In this paper, we demonstrate the studies using the LIGKA, HAGIS, HMGC and TRIMEG codes based on different levels of approximations.  
\subsection{Physical models}
Following our previous theoretical and numerical work \cite{Lu-SB-18-NF,meng2020effects}, we focus on the effects of MSSB on EP transport. As a new aspect in this work we consider more realistic AE mode structures for the further analysis in order to assess the importance of the MSB effects in present-day experiments.
In the complex Gaussian expression $A(s)=e^{- \sigma (s-s_0)^2}$ of radial envelope, 
the MSSB parameters $\sigma$ and $s_0$ are obtained by fitting the radial mode structures simulated by the gyrokinetic eigenvalue code LIGKA \cite{lauber2007ligka}, the initial value hybrid MHD-gyrokinetic code HMGC \cite{Briguglio-95-HMGC,HMGC1998} and the full-f gyrokinetic code TRIMEG-GKX \cite{lu19trimeg,lu2021trimeg-gkx}. With the more consistent $\sigma, s_0$ as the input of HAGIS \cite{pinches1998hagis} code, the EP transport is studied. The four codes employed in this work is briefly described as follows. We summarize the important properties of the employed codes in table \ref{tab:codes}.

\subsubsection{LIGKA and HAGIS}
The	LIGKA code \cite{lauber2007ligka} has been originally developed as a non-perturbative, linear	gyrokinetic	eigenvalue solver. The quasi-neutrality equation and the gyrokinetic moment equation together with the gyrokinetic equation for the particle distribution functions form a consistent model and are solved for electromagnetic perturbations in tokamak geometry \cite{qin1999linear}. It can use the pre-calculated orbits from the HAGIS code to integrate the kinetic equations for all species (electrons, ions and fast ions). Finite Larmor radius (FLR) effect and the Finite Orbit Width (FOW) effect are taken into account up to 4-th order \cite{lauber2018analytical}, consistent with the small parameter $k_\perp\rho_{t,EP}\ll1$, where $\rho_{t,EP}$ is the EP Larmor radius. For this work, the analytical coefficients as given in  \cite{lauber2018analytical} are used, and the non-linear eigenvalue problem is solved using an inverse vector iteration method \cite{Lauber2013_PREP}.

HAGIS \cite{pinches1998hagis} is an initial value particle code. HAGIS uses the linear AE information as given by LIGKA. The implementation in the ITER Integrated Modelling \& Analysis Suite (IMAS) framework \cite{imbeaux2015design} enables the convenient access to the experimental data and the interface with other codes.
In HAGIS, particles are pushed according to the guiding centre's equation of motion, and the wave-particle nonlinearity is modelled via the Lagrangian equation of the wave-particle system. The non-perturbative mode structures, frequencies and damping rates can be taken from the simulation results of LIGKA code, HMGC code or TRIMEG-GKX code.

\subsubsection{Hybrid code HMGC}
The code HMGC \cite{Briguglio-95-HMGC,vlad1995linear} is a hybrid MHD-gyrokinetic code. It describes the bulk plasma by a set of nonlinear reduced MHD equations expanded to the third order of inverse aspect ratio O($\epsilon^3$) \cite{izzo1983effects}, where $\epsilon\equiv a/R_0$ is the inverse aspect ratio, $a$ and $R_0$ are the minor and major radius, respectively. Thus, the HMGC adopts a equilibrium with circular shifted magnetic surfaces, also assuming zero bulk plasma pressure, $T_e=T_i=0$. This is the case we consider in this work, although there is an extended version of the HMGC which can treat thermal electrons as massless fluid and thermal ions as a driftkinetic species \cite{wang2011extended}. The EPs are described by the nonlinear driftkinetic equations, and coupled in the momentum equation of the bulk plasma \cite{park1992three} via the divergence of the pressure tensor term of the EP species.
Due to the underlying O($\epsilon^3$)-reduced MHD equations, the reference equilibrium analyzed by HMGC is reshaped to a circular one with $\epsilon=0.1$.

\subsubsection{Gyrokinetic full-f code TRIMEG-GKX}
TRIMEG is a TRIangular MEsh based Gyrokinetic code with multiple species. It's originally developed using $\delta f$ method and the electrostatic kinetic model, using particle-in-cell in the poloidal plane and a particle-in-Fourier scheme in the toroidal direction \cite{lu19trimeg}. It has been extended to handle electromagnetic kinetic problems including kinetic electrons using a full-f method, named TRIMEG-GKX \cite{lu2021trimeg-gkx}. The implicit particle-field solver has been developed in order to treat the fast parallel motion and large accelerations of electrons in the ``symplectic ($v_\parallel$)'' formula, and to avoid the numerical issues which corresponds to the ``cancellation problem" in the ``$p_\parallel$" formula.
The finite element method is applied in radial direction and the Fourier decomposition is used in the toroidal and poloidal direction. 
Correspondingly, the particle-in-cell is used in the radial direction and the particle-in-Fourier in the poloidal and toroidal directions. A certain amount of Fourier harmonics are kept in poloidal and toroidal directions. In this work, we keep the $n=2$ toroidal harmonic and the poloidal harmonics in the range $m\in[2,6]$ for RSAE simulation.
 
\begin{table}[htbp]\centering 
	\caption{\label{tab:codes} Comparison of simulation models used in this benchmark.}
	\footnotesize
	\begin{tabular}{l|l|l|l|l|l}
	\hline
Code & Electrons & Ions & EP & type & scheme \\ \hline
LIGKA\cite{lauber2007ligka} & Kinetic   & Kinetic                 & Kinetic &  Eigenvalue & -- \\ \hline 
HAGIS\cite{pinches1998hagis} & No       & No                      &  driftkinetic    &  Initial value & $\delta f$ \\ \hline 
HMGC\cite{Briguglio-95-HMGC}  & \multicolumn{2}{c|}{Single MHD fluid } & PIC gyrokinetic & Initial value &  $\delta f$ \\ \hline 
TRIMEG-GKX\cite{lu19trimeg,lu2021trimeg-gkx} & PIC GK  & Only $\delta n_i$ (polarization)  & PIC GK  & Initial value & Full f  \\  \hline

	\end{tabular}
\end{table}

\subsection{Parameters}
We use a circular equilibrium with Grad–Shafranov shift with parameters matched to ASDEX-Upgrade in the same way as in previous studies \cite{meng2020effects} with LIGKA and HAGIS. We use flat density and temperature profiles for both electrons and ions, i.e., $n_e = n_i = 1.71 \times 10^{19} \; m^{-3}$, $T_e = T_i = 2 \; keV$.
The equilibrium used in HAGIS and LIGKA is the same as in our previous studies \cite{meng2020effects}. As radial coordinate we use $s=\sqrt{\psi_{pol}/\psi_{pol,edge}}$, where $\psi_{pol}$ is the poloidal flux and $\psi_{pol,edge}$ is the poloidal flux at plasma edge.
When converting the flux surface coordinate $s$ to the normalized radius, we use the normalized radial coordinate $\bm{r}=r/a$, where $r$ is the radius of flux surfaces, and $a=0.4696\;m$ is the radius of last closed flux surface. 
Other parameters are as follows: major radius $R_0=1.666\;m$, magnetic field strength and safety factor at axis are $B_0=2.208\;T$. The \alfven velocity is $v_A=8.265\times 10^6\;m/s$.
The minimum of $q$ is $q_{min}=1.903$ at $s_{min}=0.494$ while $(r/a)_{min}=0.451$. 
(HAGIS and LIGKA use the same circular equilibrium with Grad-Shafranov shift. Radial coordinates $s=\sqrt{\psi_p}$.)
For HMGC, we use the nominal $a$ but an inverse aspect ratio $\epsilon=a/R_0 =0.1$.
TRIMEG-GKX uses concentric circular magnetic flux surfaces.
The $q$ profiles used in different codes are shown in Fig. \ref{fig:mode2d}, together with a typical RSAE mode structure. The $q$ profiles used in all codes agree well for $r\in [0.1a, 0.8a]$ with $q_{min}=1.903$ at $r/a=0.451$.

\section{Simulations and analyses}\label{sec3}

\subsection{Linear results w/o EP benchmarked by LIGKA, HMGC, TRIMEG-GKX}
The simulations of RSAE without EPs are benchmarked using LIGKA, HMGC, and TRIMEG-GKX. In HMGC, an antenna is used to excite the weakly damped ideal MHD RSAE in the simulations. In TRIMEG-GKX, a density perturbation is initialized and the simulation runs until clear and steady mode structure is generated and the mode frequency is measured by fitting the wave phase. The simulation is with only kinetic electron and the ion response is described with the polarization density perturbation. As described above, the LIKGA employs an inverse vector iteration method to calculate the mode structure and its complex frequency.

\begin{figure}[htbp]
{\includegraphics[width=8cm]{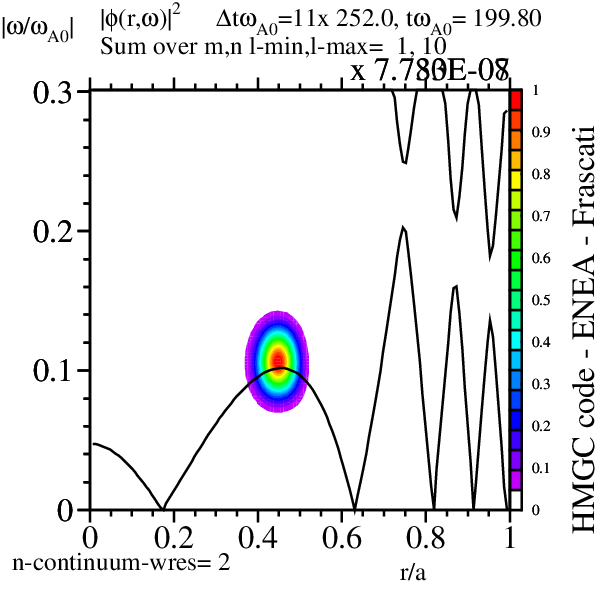}}
{\includegraphics[width=10cm]{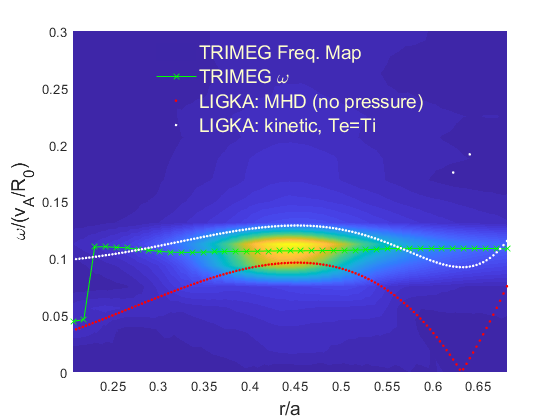}}
\caption{Left: Antenna excitation of $n=2$, $m=4$ RSAE using HMGC. Alfv\'en continua and frequency spectrum w/o thermal ion pressure. Right: spectrum of the scalar potential in the $(r,\omega)$ plane obtained from TRIMEG-GKX simulation (with kinetic electron and polarization ion density perturbation).
The green cross dots are the fitted mode frequency. The shear Alfv\'en continuum of $n=2$ are calculated by LIGKA.
The same continuum for MHD (no pressure) and that including the thermal ions and electrons kinetic effect ($T_e=T_i$)  are shown by red and white dots, respectively. 
}
\label{fig:HMGC_TRIMEGspc}
\end{figure}

\begin{figure}[htbp]
{\includegraphics[width=5.6cm]{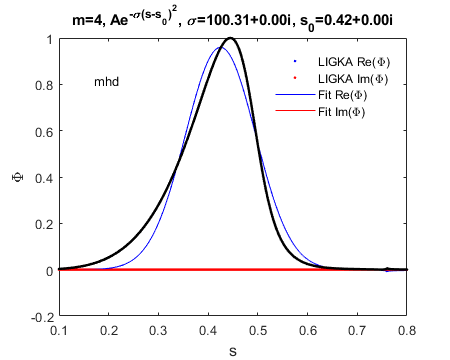}}
{\includegraphics[width=5.6cm]{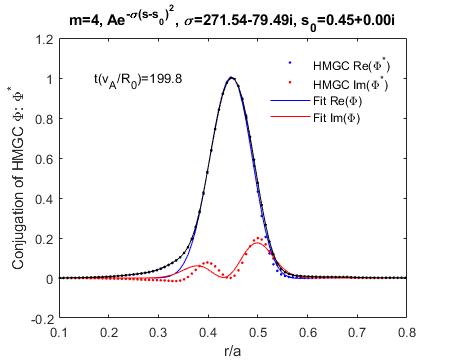}}
{\includegraphics[width=5.6cm]{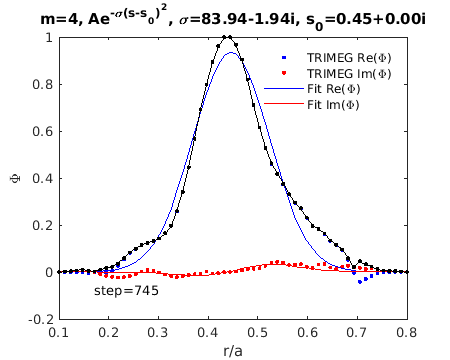}}
\caption{Mode structures from LIGKA, HMGC and TRIMEG-GKX without EPs applied.}
\label{fig:modeWOEP}
\end{figure}

The left frame of Fig. \ref{fig:HMGC_TRIMEGspc} shows the HMGC results of antenna excitation of the RSAE. The \alfven continuum and frequency spectrum are calculated w/o thermal ion pressure.
On the right frame of Fig. \ref{fig:HMGC_TRIMEGspc}, the color map of the spectrum of the scalar potential in the $(r,\omega)$ plane is from TRIMEG-GKX simulation. The green cross dots are the fitted mode frequency. The shear Alfv\'en continua of $n=2$ are calculated by LIGKA. The red dot line is the $m=4$ branch in the reduced MHD limit and the white dot line is up-shifted due to the inclusion kinetic effects related to the finite background pressure. 
Next, the frequencies and mode structures from LIGKA, HMGC and TRIMEG-GKX are compared.
As shown in table \ref{tab:modefreq},  the mode frequency agrees very well.
The imaginary part of the mode structure's radial profile is negligible (LIGKA and TRIMEG-GKX) or much smaller than (HMGC) the real part, indicating the relatively (compared with the EP driven RSAE in the next section) up-down symmetric 2D mode structure when no EPs are applied, as shown in Fig. \ref{fig:modeWOEP}. 

\begin{table}[htbp]\centering 
	\caption{\label{tab:modefreq} Mode frequency.}
	\footnotesize
	\begin{tabular}{l|c|c|c}
	\hline
		    &  LIGKA (MHD)     & HMGC (Antenna excitation no EP)  &   TRIMEG-GKX (Only kinetic electrons) \\ \hline
	$\omega$ (unit: $v_A/R_0$)  & 0.1066  & 0.1065 &  0.1049 \\ 
	\hline
	\end{tabular}
\end{table}

\subsection{EP profiles for different $n_{EP}$ curvatures at $q_{min}$}
\begin{figure}[htbp]
\centerline{\includegraphics[width=8cm]{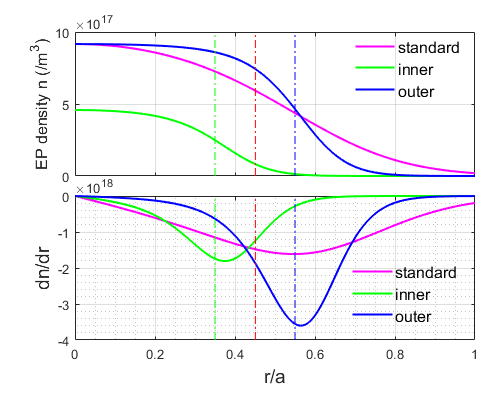}}
\caption{EP density profiles for standard, inner and outer drive cases.}
\label{fig:n_EP}
\end{figure}

In the following, EP driven RSAEs are simulated. 
A Maxwellian EP distribution with $E_0=40 \; keV$ is used. 
Different EP radial density profiles are chosen to investigate the non-perturbative EP effects.
Three typical $n_{EP}$ profiles referred to as ``inner'', ``standard'' and ``outer'' cases are used in the simulation as shown in Fig. \ref{fig:n_EP}.
The EP radial density profile is
\begin{equation}
f(r)=\frac{1}{1+\exp(\frac{r^2-r_0^2}{\delta r^2})},
\end{equation} 
where $r_0$ and $\delta r$ indicate the center and the width of the EPs' radial profile. The EP density at magnetic axis of the standard and outer cases is $n_{EP,0}=9.163\times 10^{17}\; m^{-3}$. The on-axis parameters are: $v_{EP} /v_A=0.1686$, $\rho_{t,EP}=1.3114\;cm$, and $n_{EP,0}/n_i \approx 0.05$, where $v_{EP}=\sqrt{E_0/m_{EP}}$ is the EP thermal speed and $\rho_{t,EP}=mv_{EP}/eB$ is the gyro-radius. 
The EP gradient is relatively localized for the inner and outer cases; the EP profile of the standard case with a wider full width at half maximum (FWHM) is the same as in our previous study \cite{meng2020effects}.
The gradient of the density, $dn_{EP}/dr$ at $q_{min}$ is similar for the inner, standard and outer cases;
the second derivative of EP density, $d^2n_{EP}/dr^2$ has an opposite sign at $q_{min}$ for inner and outer cases. 
By adopting these EP radial profiles, the EP drive is shifted away from $q_{min}$ and $\Delta r=r_0-r_{q_{min}}$ indicates the distance from EP drive to $q_{min}$ as shown in table \ref{tab:n_EP}. The density at axis $n_{EP,0}$ of the inner case is $1/2$ of the standard and outer cases, then the linear growth rate is similar for three cases.
\begin{table}[htbp]\centering 
	\caption{\label{tab:n_EP} EP drive distance of inner and outer cases.}
	\footnotesize
	\begin{tabular}{c|c|c}
	\hline
		       &Inner $r_0=0.35$ & Outer $r_0=0.55$\\ \hline
	      $\Delta r$ & $-0.1$  & $0.1$     \\ \hline
	      $\delta r^2$ & $0.05$ & $0.072$  \\\hline
	\end{tabular}
\end{table}

\subsection{Mode structure symmetry breaking due to non-perturbative EP effects}
The workflow of hierarchical studies is introduced in this section, by making use of the different codes, namely, LIGKA, HMGC and TRIMEG-GKX. The purpose of this hierarchical workflow is to make maximum use of physics models with different levels of fidelity while make the simulation computationally affordable. 
Firstly we calculate the mode structures with LIGKA, HMGC and TRIMEG-GKX simulations; then we fit the mode structure to get the parameters $\sigma$ and $s_0$; then substitute the more consistent $\sigma$, $s_0$ and the corresponding EP profiles in HAGIS, and study the effects $\sigma$, $s_0$ on EP transport, especially on $u_\parallel$. More self-consistent simulation of EP transport in a gyrokinetic full-f code requires much more expensive computational resources and is beyond the scope of this work. 

\begin{figure}[htbp]
\centerline{\includegraphics[width=11cm]{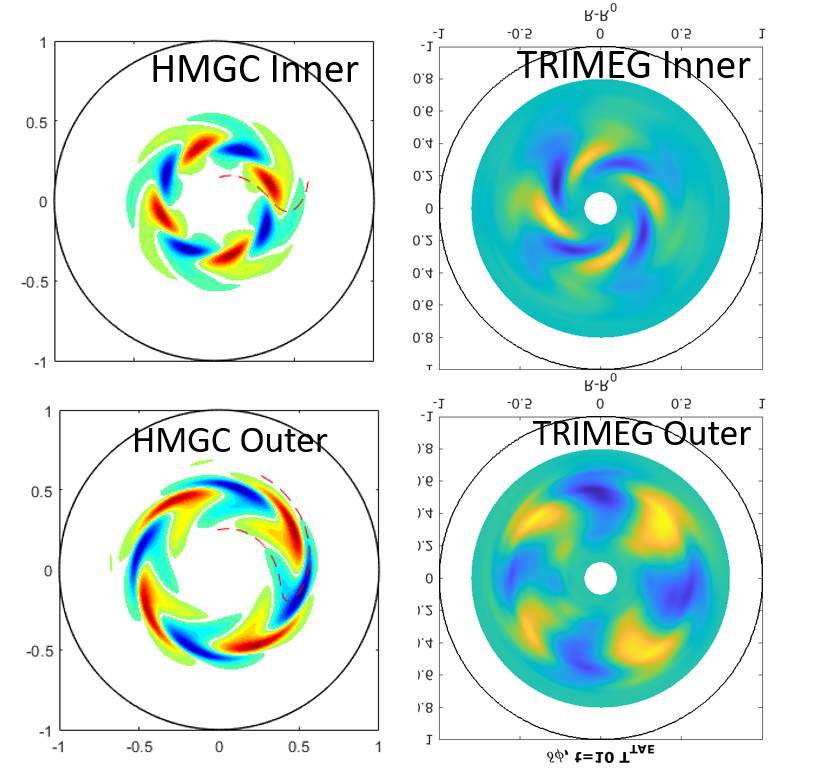}}
\caption{Mode structure symmetry breaking due to EP non-perturbative effects.}
\label{fig:mode2d}
\end{figure}

The 2D mode structures for inner and outer cases are compared as shown in Fig. \ref{fig:mode2d}.
The mode is located near the $q_{min}$ and the frequency is close to the \alfven continuum. The dominant poloidal harmonic is $m=4$.
In the MHD limit, the \alfven mode structure is up–down symmetric. EPs break this symmetry and lead to distorted mode structures (symmetry breaking).
Due to the different orientation of the poloidal angle of HMGC and TRIMEG-GKX, the 2D mode structures of the electrostatic potential perturbation $\Re(\delta\phi)$ of TRIMEG-GKX results are flipped over vertically to facilitate the comparison. As shown in Fig. \ref{fig:mode2d}, the radial wave-front of the mode has similar curvature compared with the results of HMGC and TRIMEG-GKX.
The two initial value codes both show the clear and steady clockwise mode rotation at this moment. It indicates the wave has a radial propagation from the wave-packet center towards outside in both inward and outward radial directions. 

Since $\delta f$ scheme is used in HMGC and the mode grows starting from a very small magnitude, the mode structure is clear in the whole linear stage before saturation. 
For TRIMEG-GKX code, the initial density perturbation is set so that the corresponding perturbed electric potential $\delta\phi$ contains $m=2-6$ poloidal harmonics with certain notable magnitude. 
Due to the EP excitation, the RSAE grows with the dominant $m=4$ component, while the relative magnitude of other harmonics get smaller. 
Since the full-f scheme is adopted in TRIMEG-GKX, the linear growth stage is much shorter than that in HMGC. As the $m=3$ perturbation is damped to the noise level, the relative magnitude of $m=3$ harmonic is larger than that in HMGC. Nevertheless, the $m=4$ component is always dominant during the linear and nonlinear saturation stage in TRIMEG-GKX, which is consistent with the HMGC and LIGKA results. In the following analyses, we will focus on the dominant harmonic $m=4$.

\begin{figure}[htbp]
\centerline{ {\includegraphics[width=6cm]{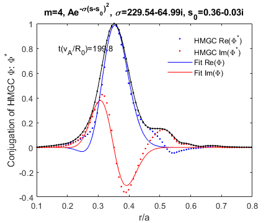}}
             {\includegraphics[width=6cm]{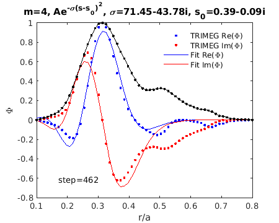}} }
\centerline{ {\includegraphics[width=6cm]{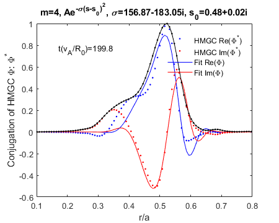}}
             {\includegraphics[width=6cm]{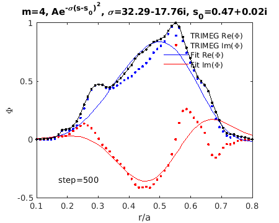}} }
\caption{Mode structure of $m=4$ harmonic of linear stage from HMGC and TRIMEG-GKX.}
\label{fig:fitmodeWEP}
\end{figure}

The more consistent parameters $\sigma,s_0$ which account for the symmetry breaking induced by the non-perturbative EP effects are determined by fitting the the radial mode structures of linear stage obtained from HMGC and TRIMEG-GKX.
As shown in Fig. \ref{fig:fitmodeWEP}, the complex Gaussian expression, $\hat{\Phi}_{nm}(s)=e^{- \sigma {(s-s_0)^2}}$, provides a good description of the $m=4$ mode structure. The mode frequency of HMGC results as shown in table \ref{tab:modefreqWEP} varies of order $10\%$ compared with no-EP case. 
The parameters $\sigma$ and $s_0$ capture the dominant features of various mode structures as shown in Fig. \ref{fig:fitmodeWEP}. The mode structure depends strongly on the location of the strongest EP density gradients due to the EP non-perturbative effects for the inner and outer cases. 

\begin{table}[htbp]\centering 
	\caption{\label{tab:modefreqWEP} Mode frequency of HMGC results.}
	\footnotesize
	\begin{tabular}{c|c|c|c}
	\hline
		                       & Inner     &  Standard  &  Outer    \\ \hline
	$\omega$ (unit: $v_A/R_0$) & 0.096     &  0.10126   & 0.0975    \\
	\hline
	\end{tabular}
\end{table}

\begin{figure}[htbp]
\centerline{\includegraphics[width=11cm]{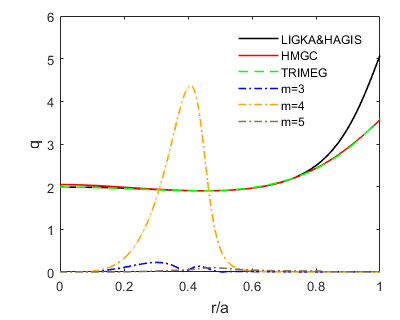}}
\caption{A typical mode structure and q profiles. The $q$ profile agrees well for $r\in [0.1a, 0.8a]$ with $q_{min}=1.903$ at $r/a=0.451$.}
\label{fig:qprofile}
\end{figure}

\begin{figure}[htbp]
\centerline{\includegraphics[width=15cm]{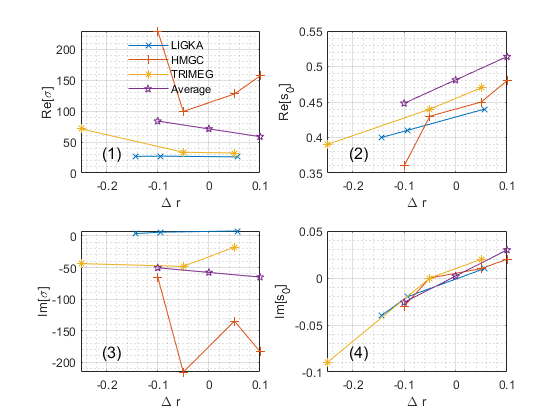}}
\caption{$\Delta r$: EP drive distance from $q_{min}$. Averaged $Re[s_0]$ is up shifted due to the difference of radial coordinates $s$ and $r/a$.}
\label{fig:sigma3codes}
\end{figure}

\begin{figure}[htbp]
\centerline{\includegraphics[width=11cm]{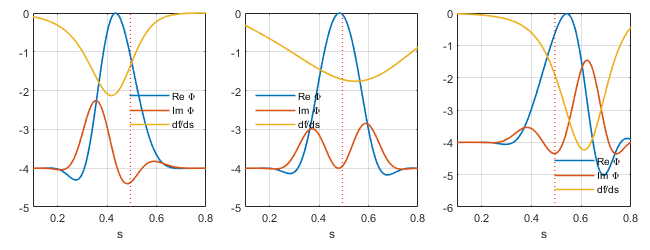}}
\caption{The mode radial structures with their corresponding EP drives for inner (left), standard (center) and outer (right) cases. The vertical red dash line shows the minimum $q$ location ($s_{q_{min}}=0.494$)}.
\label{fig:avgmodes}
\end{figure}

The $\sigma$ and $s_0$ with respect to $\Delta r$, the EP drive deviation from the $q_{min}$, is shown in Fig. \ref{fig:sigma3codes}.
The radial width of the mode structure is proportional to $\sqrt{1/\Re{(\sigma)}}$. The FWHM of the mode normalized to minor radius, $W_m=2\sqrt{ln2/\Re(\sigma)}$, is $0.1665$ when $\Re(\sigma)=100$.
The $\Re(\sigma)$ of HMGC is relatively larger, i.e., the mode is narrower. This is because of the aspect ratio of 10 and the slightly different $q$ profile used in HMGC as shown in Fig. \ref{fig:qprofile}.
The $\Re(s_0)$ dominantly defines the mode peak location. As shown in Fig. \ref{fig:sigma3codes}(2) $\Re(s_0)$ tends to follow the EP drive center. As shown in Fig. \ref{fig:sigma3codes}(3),  $\Im(\sigma)<0$  for inner, standard and outer cases of HMGC and TRIMEG-GKX results. It is consistent with that HMGC and TRIMEG-GKX both show that the wave propagates from the wave-packet center towards two sides in radial direction. 
The $\Im(\sigma)$ of LIGKA results is relatively small, which means that the mode propagates in the radial direction mainly in one direction. The origin of this difference is related to the eigenvalue approach: due to the flat background profiles, the radial eigenstates are almost degenerated in the frequency domain. That means, various RSAEs with a different radial mode numbers can be found for this case. Whereas in an initial value code these states mix and form an overall linear mode structure, in an eigenvalue code they all appear separately, but with very similar frequency and damping/growth rates. Multiple eigenstates corresponding to radial node numbers can leads to the difference between LIGKA and initial value codes TRIMEG-GKX and HMGC. The $\Im(s_0)$ value shows similar trends as HMGC/TRIMEG, which is the key of EP $u_\parallel$ reversal, while the $\Im(\sigma)$ does not affect $u_\parallel$ reversal too much \cite{meng2020effects}.
The $\Im(s_0)$ for the results of the three codes changes sign for inner and outer cases as shown in Fig. \ref{fig:sigma3codes}(3). It means the averaged radial wave vector $\langle k_s \rangle$ changes from negative to positive. Here, the averaged radial wave vector is defined as $\langle k_s \rangle =\int \Re(k_s) \cdot |\Phi|^2 ds / \int |\Phi|^2 ds$. The purple lines are the averaged $\sigma$ and $s_0$ as shown in Fig. \ref{fig:sigma3codes}. We fit a linear function to the results of each code and evaluate the $\sigma$ and $s_0$ at points $\Delta r=-0.1,0,0.1$ by the polynomial coefficients. Then we average the three fitted $\sigma$ and $s_0$ at $\Delta r=-0.1,0,0.1$, respectively. Note that the averaged $Re[s_0]$ shown in Fig. \ref{fig:sigma3codes}(3) is up shifted by $0.043$ due to the difference of radial coordinates $s$ and $r/a$, $s_{q_{min}}-(r/a)_{q_{min}}=0.043$. 
Using the average parameters $(\sigma,s_0)$ from the three codes, the fitted mode structures are constructed for the inner, standard and outer cases, and shown together with the EP density gradients in Fig. \ref{fig:avgmodes}. 
The mode location follows the EP drive and is between the maximum EP density gradient and the location of $q_{min}$ for inner and outer cases.
These are used as the input of HAGIS to study the EP transport next.

\subsection{EP transport with the effects of mode structure symmetry breaking}
 
\begin{figure}[htbp]
\centerline{\includegraphics[width=11cm]{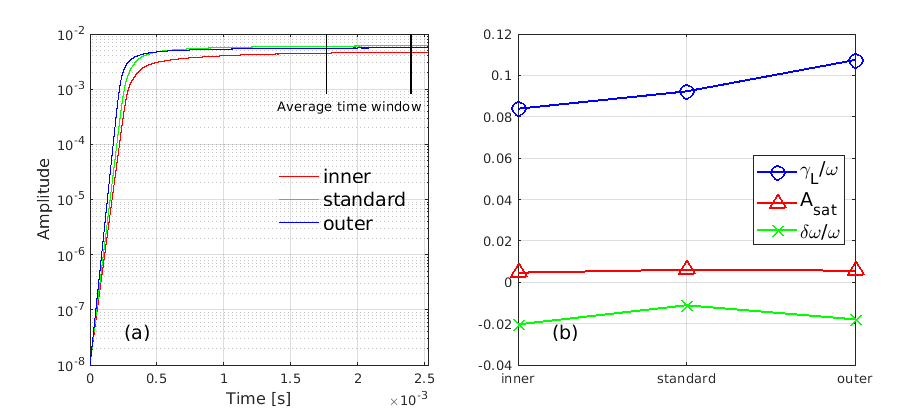}}
\caption{(a): Mode evolution of 3 cases. (b): The comparison of linear growth rate $\gamma_L$, the frequency shift $\delta\omega$ and the saturation level $A_{sat}$.}
\label{fig:growsat}
\end{figure}

\begin{figure}[htbp]
{\includegraphics[width=5.6cm]{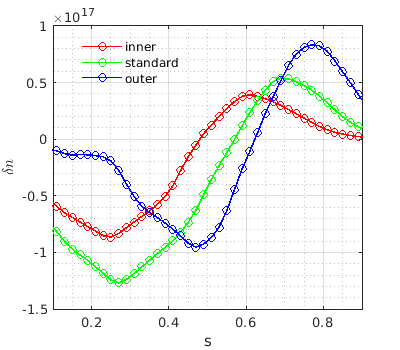}}
{\includegraphics[width=5.6cm]{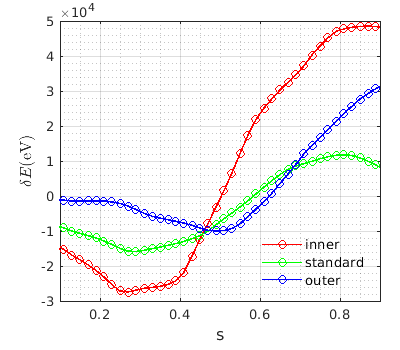}}
{\includegraphics[width=5.6cm]{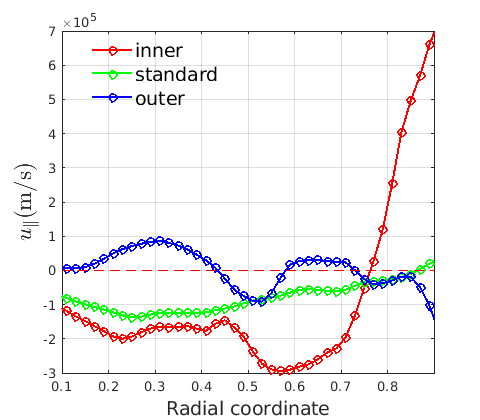}}
\caption{The radial profiles of the perturbed EP density (left), energy  (middle) and the parallel velocity (right). The integration is performed over velocity space and other configuration space coordinates in obtaining the radial profile.}
\label{fig:delta_f}
\end{figure}

The EP transport is studied with the consideration of the MSSB. The EP profiles and the more consistent, fitted mode structures from LIGKA, HMGC and TRIMEG-GKX are adopted as the inputs of HAGIS. The eigenmode frequency $\omega=0.1066(v_A/R_0)$, i.e., $f=79.2 \; kHz$, is used for inner, standard and outer cases. 
The time evolution of the mode amplitude, mode phase and particles are solved using HAGIS. 
For the outer, standard and inner cases, the saturation level $A_{sat}$ and EP induced linear growth rate $\gamma_L$ and frequency shift $\delta\omega$ are similar, as shown in Fig. \ref{fig:growsat}. 
The saturation level $A_{sat}$ and the frequency shift $\delta\omega$ in Fig. \ref{fig:growsat}(b) are averaged values over the time window as shown in Fig. \ref{fig:growsat}(a). The mode has a similar saturation level such that the EP transport is studied at similar wave intensity levels. The EP density and energy profile flattening occurs at the mode location as shown in Fig. \ref{fig:delta_f}. Especially, the momentum transport depends significantly on the initial EP profile. As shown in the right frame of Fig. \ref{fig:delta_f}, as the EP density gradient shifts from the inner to the outer position, $u_\parallel$ changes from negative to positive, with a magnitude of order of $100\; km/s$. With stronger EP drive in this more consistent simulation, the magnitude of $u_\parallel$ is larger than the previously investigated weaker EP drive case \cite{meng2020effects}.

\begin{figure}[htbp]

\centerline{\includegraphics[width=11cm]{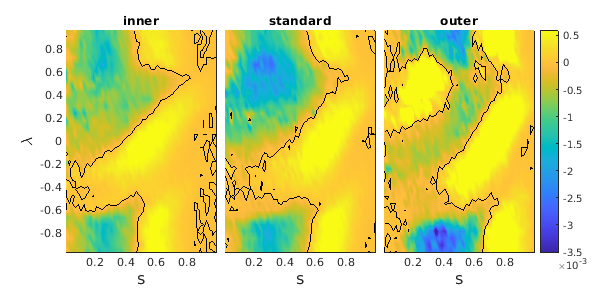}}
\caption{$\delta f(\lambda,s)$ of EPs for inner, standard and outer cases, where $\lambda=v_\parallel/v$. The black line is the contour of $\delta f=0$.}
\label{fig:phasespace_deltaf}
\end{figure}

The perturbed distribution function $\delta f(\lambda,s)$ in radial and pitch angle is shown in Fig. \ref{fig:phasespace_deltaf}, where $\lambda=v_\parallel/v$.
The change of the $\delta f$ value is more visible for the co-moving particles, indicating the close connection between $u_\parallel$ reversal and the phase space structure of the wave-particle interaction. For the outer case, the co-moving particles interact with the wave strongly and contribute to the generation of the positive $u_\parallel$ in the inner ($s<0.4$) region.
While the asymmetric particle
response of thermal ions to the microturbulence has been studied in previous work \cite{wang2010nonlinear,peeters2011overview}, our results demonstrate that for EP driven RSAEs, the asymmetric mode structure also plays an
important role in particle response and the $u_\parallel$ generation.

\section{Summary and conclusion}\label{sec4}
In this work, hierarchical simulations of RSAE with or without EP have been performed using the gyrokinetic eigenvalue code LIGKA, the initial value gyrokinetic $\delta f$ particle code HAGIS, the hybrid MHD gyrokinetic code HMGC and the full-f gyrokinetic particle code TRIMEG-GKX.
Self-consistent simulations of EP driven Reversed Shear \alfven Eigenmode demonstrate the MSSB due to the EP effects. 
It is shown that the mode structure and frequency agree well between LIGKA, HMGC and TRIMEG-GKX and the mode structure is more symmetric when EPs are not applied. 
The analytical complex Gaussian expression $\exp\{- \sigma (s-s_0)^2\}$ provides a good description of the mode structure in linear stage and with properly fitted parameters $\sigma$ and $s_0$, it captures the dominant features of various mode structures with symmetry breaking properties.
The modes symmetry breaking has a non-ignorable $\Im(\sigma)$ when EP are included for typical parameters in ASDEX-Upgrade discharges \cite{lauber2018IAEA};
the $\Im(s_0)$ has opposite signs when the EP drive moves from one side of $q_{min}$ to the other side.
The $u_\parallel$ reversal is observed as EP initial profiles varies in HAGIS simulation, when the more consistent RSAE structures are adopted. 
Thus, our result shows that the EP-RSAE interaction induced $u_\parallel$ changes direction when the EP drive moves from one side to the other side of $q_{min}$ for the parameters studied in this work.

While in this work, the effects of the mode structure distortion on EP transport is demonstrated, further work will be done in future by using more realistic parameters, simulation models and developing the EP transport model for interpretive and predictive studies. With more realistic equilibrium and profiles used, the simulation is expected to be more quantitatively accurate in EP transport for non-circular tokamak plasmas. While the turbulence induced transport has been studied using gyrokinetic simulations \cite{wang2010nonlinear,peeters2011overview}, the gyrokinetic simulations and analyses of toroidal/parallel flow generation with electromagnetic and EP physics have not carried out so far. Consistent simulations in the presence of EPs require the proper treatment of kinetic effects of Alfv\'en waves and EPs \cite{chen2016review}, consistent gyrokinetic ordering \cite{brizard2007foundations} and proper treatment of conservation properties \cite{scott10GKmomentum}. Meanwhile, the development of EP transport models serves as an extension of this kind of near-first-principle studies with the aim to support the interpretation and prediction of experimental results. \cite{meng2018NF,white2019collisionResBroad,berk95LBQ,nikolaiRBQnf,zonca2015nonlinear}. 


\section*{Acknowledgements}
The authors would like to thank Dr. F. Zonca, Dr. X. Garbet and Dr. Simon Pinches for fruitful discussions, partially within the EUROFUSION Enabling Research Projects Projects ``NLED'' (ER15-ENEA-03), ``NAT'' (CfP-AWP17-ENR-MPG-01), ``MET'' (ENR-MFE19-ENEA-05) and ``ATEP''(ENR-MOD.01.MPG). This work has been carried out within the framework of the Eurofusion Consortium and has received funding from the Euratom research and training programme 2014-2018 and 2019-2020 under grant agreement No 633053. The views and opinions expressed herein do not necessarily reflect those of the European Commission.

\bibliographystyle{unsrt}

\bibliography{references}

\end{document}